\documentclass[twocolumn]{jpsj2} 
\usepackage{amsmath}

\title{
Absence of Hybridization Gap in Heavy Electron Systems and Analysis of YbAl$_3$ in terms of Nearly Free Electron Conduction Band
}

\author{%
Hiroyuki {\sc Kuroiwa}, Yoshiki {\sc Imai} and Tetsuro {\sc Saso}
}

\inst{%
Department of Physics, Saitama University, Saitama, 338-8570
}

\recdate{\today}

\abst{
In the analysis of the heavy electron systems, theoretical models with c-f hybridization gap are often used.  
We point out that such a gap does not exist and the simple picture with the hybridization gap is misleading in the metallic systems, and present a correct picture by explicitly constructing an effective band model of YbAl$_3$. Hamiltonian consists of a nearly free electron model for conduction bands which hybridize with localized f-electrons, and includes only a few parameters. 
Density of states, Sommerfeld coefficient, f-electron number and optical conductivity are calculated and compared with the band calculations and the experiments.
}

\kword{
YbAl$_3$, nearly free electron model, empty-core pseudo-potential
}
\begin{document}
\maketitle
%
\section{Introduction}
\begin{figure}[tb]
\begin{center}
\includegraphics[width=8.5cm]{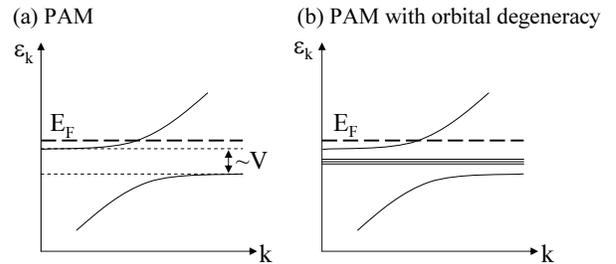}
\end{center}
\caption{Schematic band dispersions (a) usual PAM and (b) PAM with orbital degeneracy of f-electrons. Dashed line stands for the Fermi level. 
}
\label{pam}
\end{figure}

The 4f-electron systems in rare-earths have attracted much interest since the duality between the itinerant and the localized properties of f-electrons gives rise to rich variety of phenomena.~\cite{hews93} 
In particular, Ce and Yb compounds have been extensively studied for several decades because Ce has only one 4f-electron and Yb only one 4f-hole. 
The periodic Anderson model (PAM) is often employed in order to analyze the properties of these compounds. The most simplified PAM neglects the orbital degeneracies of both conduction and f-electron states, and is given by
\begin{eqnarray}
{\cal H}&=&
 \sum_{\mathbf k\sigma}\epsilon^{\rm c}_{\mathbf k} c^{\dag}_{\mathbf k \sigma}c_{\mathbf k \sigma}
+\sum_{i\sigma} E_{\rm f} f^{\dag}_{i\sigma}f_{i\sigma} \nonumber \\ 
&+& \sum_{\mathbf k\sigma}(V_{\mathbf k \sigma}c^{\dag}_{\mathbf k \sigma}f_{\mathbf k \sigma}+{\rm h.c.} )
+U \sum_i n^{\rm f}_{i \uparrow}n^{\rm f}_{i \downarrow}
\label{eqn:pam}, 
\end{eqnarray}
where $c^{\dag}_{\mathbf k \sigma}$ ($f^{\dag}_{\mathbf k \sigma}$) is a creation operator of a conduction (f) electron with the wave vector $\mathbf k$ and the spin $\sigma$. $n^{\rm f}_{i \sigma}$ represents the local f-electron number operator (=$f^{\dag}_{i \sigma}f_{i \sigma}$) with the site index $i$. $\epsilon^{\rm c}_{\mathbf k}$ is the conduction band energy and $E_{\rm f}$ the f-electron energy level. $V_{\mathbf k\sigma}$ represents the hybridization between the conduction and f-bands (c-f hybridization) and $U$ is the Coulomb interaction between the f-electrons. 
${\mathbf k \sigma}$-dependence of $V_{\mathbf k\sigma}$ is often neglected.
This Hamiltonian is regarded as the minimal model to describe the 4f- and 5f-electron systems and has been partially successful in qualitatively capturing the physical properties. Especially, the heavy mass is explained by the combination of the position of the Fermi level intersecting the nearly flat part of the hybridized bands (see Fig. \ref{pam}(a)) and the renormalization due to strong correlations in f-electrons. 
Furthermore, a hybridization gap appears and the width is of the order of $V$ for the non-interacting case. 
Using this simple model, Okamura, {\it et al.} found a universal scaling property on the peak position and the hybridization strength over many Ce and Yb compounds.\cite{okam07}

The degeneracy of the f-states is very important. To incorporate it, the following model has been investigated in detail:\cite{yama87}
\begin{eqnarray}
{\cal H}
\hspace{-3mm}
&=&
\hspace{-3mm}
 \sum_{\mathbf k\sigma}\epsilon^{\rm c}_{\mathbf k} c^{\dag}_{\mathbf k \sigma}c_{\mathbf k \sigma}
+\sum_{iM} E_{\rm f} f^{\dag}_{iM}f_{iM} \nonumber \\ 
\hspace{-3mm}
&+& 
\hspace{-3mm}
\sum_{\mathbf k \sigma M}(V_{\mathbf k\sigma M}c^{\dag}_{\mathbf k \sigma}f_{\mathbf k M}+{\rm h.c.} )
+U 
\hspace{-3mm}
\sum_{i,M>M'} n^{\rm f}_{iM}n^{\rm f}_{iM'}
\label{eqn:pam}.
\end{eqnarray}
Here, the conduction electrons have been treated as free plane waves, and $M$ denotes the $z$-component of the total angular momentum $J$ of the f-electrons.
Anderson, however, pointed out that only two of $(2J+1)$-fold degenerate f-states have finite mixing matrix elements (see eq. (\ref{eqn:clbgld72}) later) with the plane waves, so that $(2J+1)-2$ f-states remain unhybridized and stay in the gap (see Fig. \ref{pam}(b)).~\cite{ande81}
Using this model, Kontani derived the grand Kadowaki-Woods relation~\cite{kont04}, which improves the original Kadowaki-Woods relation by taking into account the effect of the degeneracy of the f-orbitals.  Comparison with the experiments is made in detail in ref.~\ref{tsuj05}. 
In this theory, the hybridization was approximated as a constant independent of ${\mathbf k}$ or $M$.

The ${\mathbf k}$-dependence of $V_{\mathbf k\sigma M}$ was investigated $e.g.$ by Hanzawa {\it et al.}~\cite{hanz87}, Ikeda and Miyake~\cite{iked96}, etc., but the conduction electrons were always treated as free plane waves.

Completely different approach to heavy electron systems is taken by using the tight-binding model to both the conduction and f-electrons.
For example, the local density approximation (LDA) energy band of the most typical Kondo insulator YbB$_{12}$ was reproduced very well by this approach, and it was clarified that the degeneracies of both the conduction and f-electrons are essentially important in the opening of the gap in this material.~\cite{saso03} 
Anisotropy in the magnetic field-induced insulator-to-metal transition was explained~\cite{izum07} 
and correlation effect on the optical conductivity was also investigated.\cite{saso04}  Application to Ce-compounds was done by Maehira, {\it et al}.\cite{hott03}

Fermi surfaces of heavy electron systems are calculated via LDA and agreement with the de Haas van Alphen (dHvA) experiments has been achieved in several compounds.~\cite{sett07} 
LDA calculation, however, does not include dynamical many-body effect, which is of much importance in many cases. 
Recently, combination of the LDA band calculation and the dynamical mean field theory (DMFT) has been rapidly developed.~\cite{kotl06} 
Many rare-earth compounds, however, have extremely small energy scale, so that it is not easy to carry out the LDA+DMFT calculations. 

\begin{figure}[bt]
\begin{center}
\includegraphics[width=7cm]{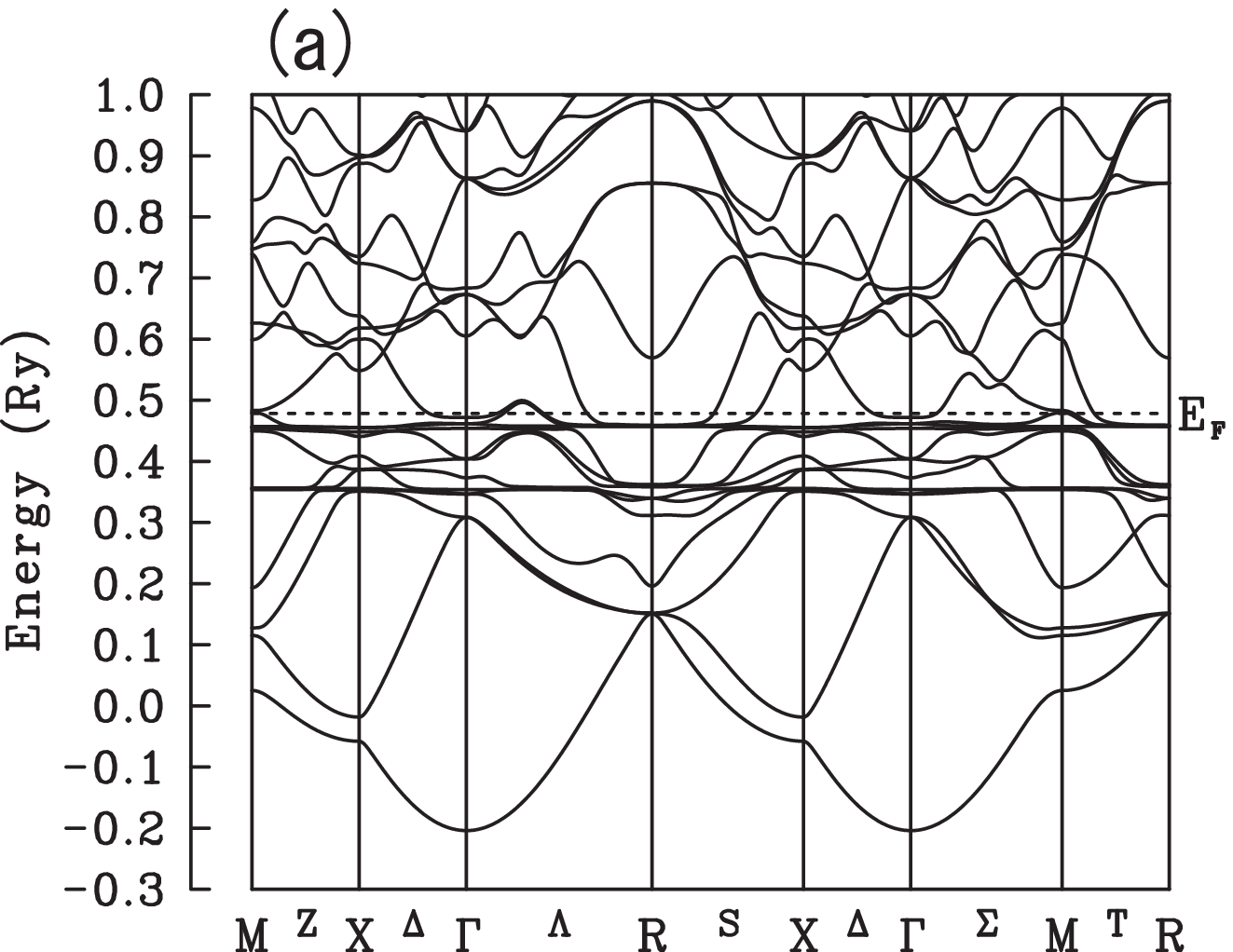}

\vspace{3mm}
\includegraphics[width=7cm]{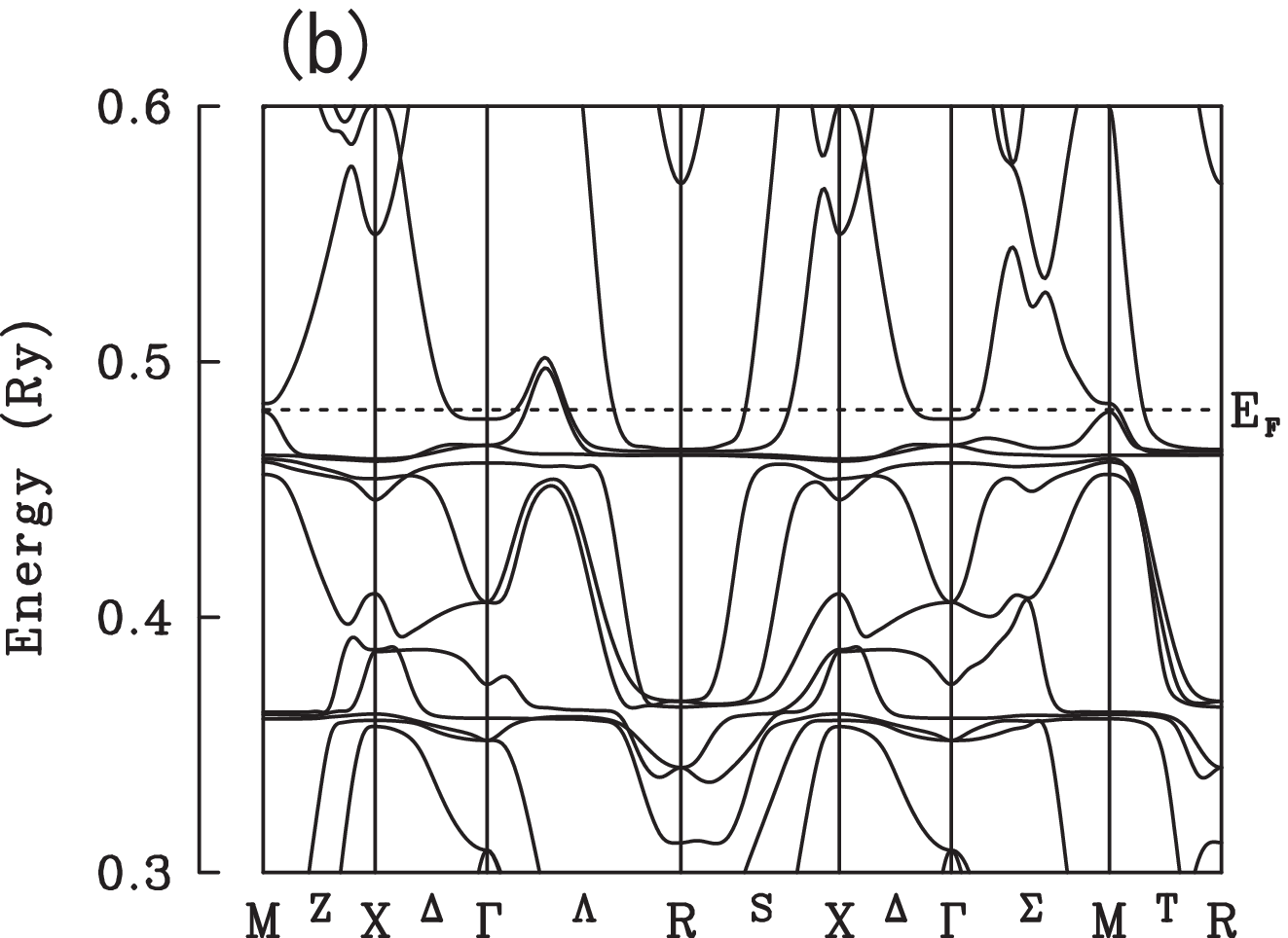}

\vspace{3mm}
\includegraphics[width=7cm]{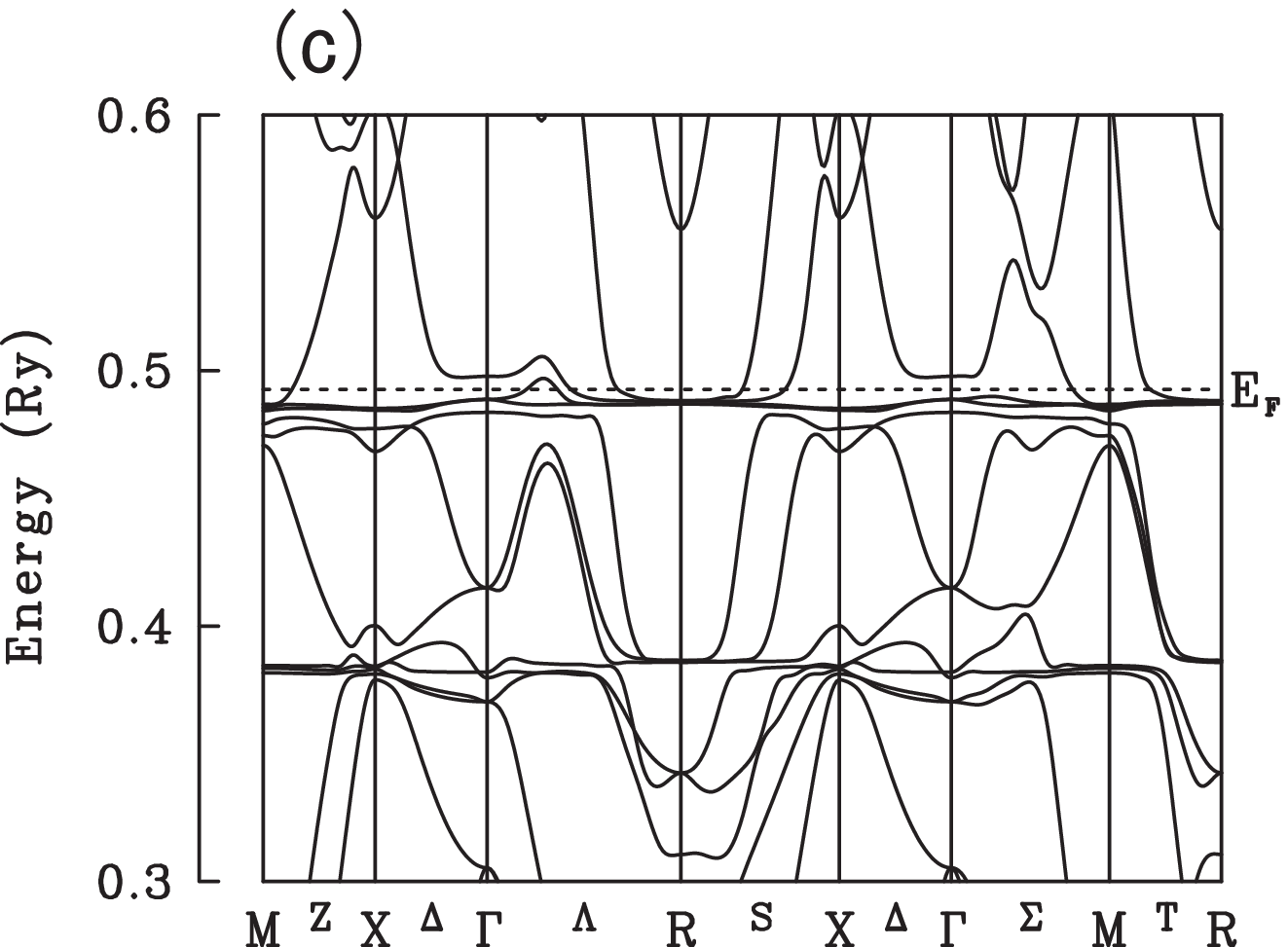}
\end{center}
\caption{LDA band structure of YbAl$_3$.~\cite{hari}
Upper panel (a) and center panel (b) show the overall band structure and its magnification near the Fermi level $E_{\rm F}$, respectively. Lower panel (c) shows the result with the upward-shift of the 4f-levels by 0.2 Ry.~\cite{ebih00} 
}
\label{ldaband}
\end{figure}

Therefore, it is desirable to construct an effective band model for the 4f-electron systems as a starting point to include strong correlations.  It is already realized by using the tight-binding model as mentioned above, but there exists other type of heavy electron material which can be more suitably described by the nearly free electron (NFE) conduction band than the tight-binding model. 
The LDA result indicates that the conduction band structure of the heavy electron metal YbAl$_3$ has the NFE character, which is shown in Fig. \ref{ldaband}.~\cite{hari} 
The band dispersions show the apparent parabolic structure in the low energy range below the Fermi energy $E_{\rm F}$. 

In this paper, we focus on the heavy electron metal YbAl$_3$ with the NFE conduction band and construct an effective Hamiltonian, 
which may become a good starting point to systematically discuss correlation effects on the physical quantities in the future. 
To take account of the effect of a weak periodic potential for the conduction electrons, we use the NFE scheme with the empty-core pseudo-potential method~\cite{harr80,harr99}, which has been successfully applied to simple metals and semiconductors. 
Furthermore, we accurately consider the hybridization effect between the NFE conduction bands and the localized f-bands, which considerably modifies the previous studies. 
Using this Hamiltonian, we calculate the band dispersions, the density of states (DOS), the Sommerfeld coefficient and the optical conductivity. 
We discuss the physical properties of typical rare-earth heavy fermion metal YbAl$_3$ by comparing with the LDA and experimental results. 

YbAl$_3$ has Cu$_3$Au-type crystal structure and belongs to the simple cubic structure as the Bravais lattice with a lattice constant 4.2 \r{A}. The effective mass observed from the dHvA measurement~\cite{ebih00} is 14 $\sim$ 24 $m_0$ where $m_0$ is the rest mass of an electron. The Sommerfeld coefficient is $\gamma\sim$ 40 mJ $\cdot$ K$^{-2}\cdot$mol$^{-1}$.~\cite{corn02,baue04} 
The mean valence is in the range from 2.65 to 2.8~\cite{tjen93,suga05}, which indicates that YbAl$_3$ is a valence-fluctuation compound. 
Note that ref. \ref{suga05} indicated the limitation of the application of the single impurity Anderson model to the periodic systems in the analysis of photoemission spectra. 
The single site Kondo temperature $T_{\rm K}$ is 500 $\sim$ 700 K~\cite{corn02}. The coherence temperature is about $T_{\rm coh} \sim 35$ K, below which the electric resistivity follows the relation: $\rho \propto T^2$.~\cite{corn02,ebih03} This result indicates explicitly that YbAl$_3$ exhibits the Fermi liquid behavior below $T_{\rm coh}$. The optical conductivity shows the pseudo-gap structure at low energy region and the mid-infrared peak at $\hbar \omega \sim$ 0.25 eV.~\cite{okam04} 

Note that the additional shift of the 4f-levels (0.2 Ry) must be introduced in the LDA calculation (Fig. \ref{ldaband}(c)) in order to reproduce the angular dependence of the dHvA frequency.~\cite{ebih00}
In comparison with Fig. \ref{ldaband}(b), the band around $\Gamma$ point near the Fermi level shifts to higher energy region, and the electron-like Fermi surface disappears. The bands around M point near the Fermi level shift down, so that the multiply-connected Fermi surface appears. 
In addition to the energy dispersion~\cite{hari}, the DOS has been obtained by LDA.~\cite{lee03} We discuss the accuracy of the constructed model by comparing with these results. 

This paper is organized as follows. We construct the effective Hamiltonian using the NFE scheme with the empty-core pseudo-potential method for the conduction bands, and introduce the hybridization with the 4f-states in the next section. In Sec. \ref{results}, we show the band dispersion, the DOS and the optical conductivity. Summary and discussions are given in Sec. \ref{sumdis}. 

%
\section{Construction of Effective model}

In this section, we construct the effective Hamiltonian for the band energy of the rare-earth compound, YbAl$_3$, which consists of three Al and one Yb atoms per unit cell. The electronic configuration of each atom is [Ne](3$s$)$^2$(3$p$)$^1$ for Al and [Xe](4$f$)$^{14}$(5$d$)$^0$(6$s$)$^2$ for Yb. 
Except the closed shell parts [Ne] and [Xe], there are 25 electrons per unit cell of YbAl$_3$. 
While it is natural to assume that Al ion becomes trivalent, the valence of each Yb ion can be divalent or trivalent, where the mean configuration of f-electrons is in the range from $f^{14}$ to $f^{13}$. It means that YbAl$_3$ is a typical valence fluctuation compound and shows a metallic state.

The conduction electrons in YbAl$_3$ consist of three valence electrons from each Al atom and 2 $\sim$ 3 valence electrons from Yb atom, which spread over in the crystal. Although the 4f-orbitals of Yb atom have the strongly localized character, the hybridization with the conduction electrons makes the f-electrons partially itinerant. Therefore, the effective Hamiltonian can be written as follows 
\begin{eqnarray}
{\cal H}^{\rm eff}&=&{\cal H}^{\rm band} +{\cal H}^{\rm int}, \\
{\cal H}^{\rm band}&=&{\cal H}^{\rm c} +{\cal H}^{\rm f} +{\cal H}^{\rm hyb}, 
\end{eqnarray}
where ${\cal H}^{\rm band}$ represents the band energy. 
${\cal H}^{\rm c}$ (${\cal H}^{\rm f}$) is the conduction (localized f) electron term and 
${\cal H}^{\rm hyb}$ is the hybridization between the conduction electrons and the f-electrons. 
${\cal H}_{\rm int}$ represents the Coulomb interaction between the f-electrons. We focus on the construction of the effective Hamiltonian for the band energy ${\cal H}^{\rm band}$ in the present study. 

Let us start with the construction of the effective Hamiltonian. First, we discuss the conduction electron term ${\cal H}^{\rm c}$. The wavefunction of conduction electrons generally spreads over on the large region in comparison with a lattice constant. Therefore, the Hamiltonian of the conduction electrons is given by
\begin{eqnarray}
{\cal H}^c=-\frac{\hbar^2}{2m}\nabla^2 +V(\mathbf r), 
\end{eqnarray}
where $V(\mathbf r)$ stands for the periodic potential term due to the ions on the lattice sites, which is written as 
\begin{eqnarray}
V(\mathbf r)=\sum_{\mathbf R} \left\{v^{\rm Yb}(\mathbf r- \mathbf R)+\sum_{\pmb{\tau}}v^{\rm Al}(\mathbf r - \mathbf R-\pmb{\tau})\right\}, 
\end{eqnarray}
where $v^{\rm Yb}$ and $v^{\rm Al}$ are atomic potentials in the Yb and Al sites. $\pmb{\tau}$ represents the coordinate of each Al ion measured from the Yb site in the unit cell, and ${\mathbf R}$ is the translation vector. 
We neglect effect of Coulomb interactions between the conduction electrons because of strong screening effects. 

The NFE method is employed for the conduction electrons in order to take account of the effect of the weak periodic potential, which is given by
\begin{eqnarray}
V(\mathbf r)=\sum_{\mathbf G} V_{\mathbf G} \, {\rm e}^{{\rm i}\mathbf G \cdot \mathbf r}. 
\label{eqn:nfevr}
\end{eqnarray}
$\mathbf G$ stands for the reciprocal lattice vector. 
$V_{\mathbf G}$ represents the matrix element between the plane waves, 
\begin{eqnarray}
V_{\mathbf G}&\equiv& \langle \mathbf k | V(\mathbf r) |\mathbf k- \mathbf G \rangle
=\frac{1}{\Omega}\int {\rm d} \mathbf r\, V(\mathbf r)\,{\rm e}^{-{\rm i}\mathbf G \cdot \mathbf r} \nonumber \\
&=&\frac{1}{\Omega_0}\int {\rm d} \mathbf r\, v^{\rm Yb}(\mathbf r)\,{\rm e}^{-{\rm i}\mathbf G \cdot \mathbf r}\nonumber \\
&+&\frac{1}{\Omega_0}\sum_{\pmb{\tau}}{\rm e}^{-{\rm i}\mathbf G \cdot \pmb{\tau}}\int {\rm d} \mathbf r\, v^{\rm Al}(\mathbf r) \, {\rm e}^{-{\rm i}\mathbf G \cdot \mathbf r}, 
\end{eqnarray}
where $\Omega$ is the volume of the system and $\Omega_0$ is the volume per unit cell.
 
Next, let us give the potential explicitly. 
From the LDA band calculation, the conduction bands below $E_{\rm F}$ look like NFE. Above $E_{\rm F}$, the contribution from d electrons deforms the bands from NFE. We particularly focus on semi-quantitative description of the conduction bands around and below $E_{\rm F}$. 
We apply the empty-core pseudo-potential method~\cite{harr80,harr99} to the description of the weak atomic potential in each site, which is given by
\begin{eqnarray}
v^{\alpha}(\mathbf r) =
\left \{
\begin{array}{cl}
0& (|\mathbf r| < r^{\alpha}_{\rm c}) \\
-\frac{Z^{\alpha}e^2}{|\mathbf r|}& (|\mathbf r| > r^{\alpha}_{\rm c}), 
\end{array}
\right. 
\end{eqnarray}
where $r^{\alpha}_{\rm c}$ represents the core radius and $Z^{\alpha}$ is the effective valence number ($\alpha$=Yb, Al), which are adjustable parameters. 

However, it is well known that aluminium is a trivalent metallic system and its band dispersion can be described well by the pseudo-potential technique. Thus, $r^{\rm Al}_{\rm c}$ and $Z^{\rm Al}$ are taken from Harrison's textbooks~\cite{harr80,harr99} in the present study. Namely, we set $r^{\rm Al}_{\rm c}=0.61$ \r{A} and $Z^{\rm Al}=+3$. 
Compared with the Al case, the estimation of Yb parameters is not so unique.  
It is because Yb ion can be divalent ($Z^{\rm Yb}=+2$) or trivalent ($Z^{\rm Yb}=+3$). 
We investigated the effects on the physical properties for both values, and obtained almost similar results. We hereafter employ $Z^{\rm Yb}=+2$. 
Furthermore, we also calculated the band dispersions for several values of the core radius $r^{\rm Yb}_{\rm c}$ in the range from 0.5 to 3.0 \r{A}. Although there was no qualitative difference for all cases, the overall band structure for $r^{\rm Yb}_{\rm c}=2.0$ \r{A} well reproduced the LDA result. Thus, we employ this value of $r^{\rm Yb}$ for a trial, although it is larger than the value in refs. \ref{harr80} and \ref{harr99}. 

The Fourier transformations of the pseudo-potentials are written as
\begin{eqnarray}
v^{\alpha}_{\mathbf q}=-\frac{4 \pi Z^{\alpha} e^2}{\Omega_0} \cdot \frac{\cos(qr^{\alpha}_{\rm c})}{q^2},
\end{eqnarray}
where we define $q \equiv |\mathbf q|$. 

\begin{figure}[tb]
\begin{center}
\includegraphics[width=7cm]{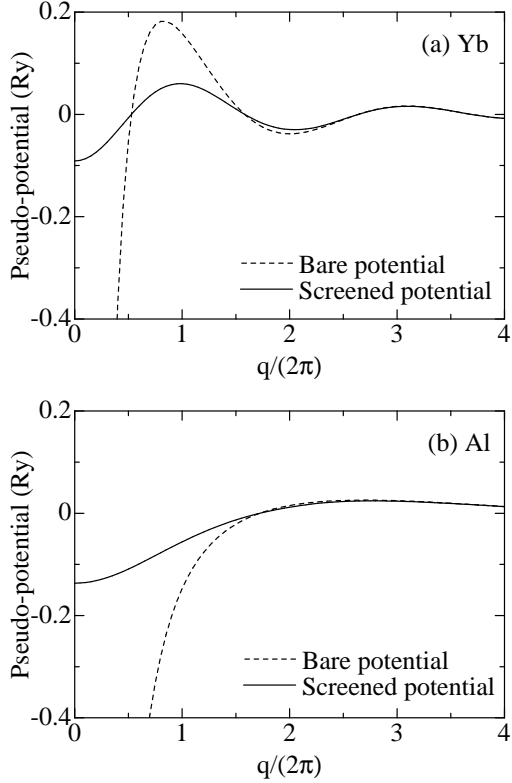}
\end{center}
\caption{
Bare and screened pseudo-potentials of Yb (upper panel) and Al (lower panel) for YbAl$_3$. The effective valence numbers are $Z^{\rm Yb}=+2$ and $Z^{\rm Al}=+3$, and the core radii are $r^{\rm Yb}_{\rm c}=2.0$ \r{A} and $r^{\rm Al}_{\rm c}=0.61$ \r{A}, respectively. }
\label{potential}
\end{figure}

The redistribution of the conduction electrons gives rise to screening of the charge at large distance. The screening effect in the pseudo-potential is considered by using the dielectric function of the free electron gas. Then, we obtain the screened potential
\begin{eqnarray}
\tilde{v}^{\alpha}_{\mathbf q}&=&\frac{v^{\alpha}_{\mathbf q}}{\epsilon ({\mathbf q})}, \\
\epsilon ({\mathbf q})&=&1+\frac{4\pi e^2}{q^2}\cdot \frac{D_{0}(E_{\rm F})}{\Omega}\cdot \Pi_0(x), 
\end{eqnarray}
where $D_{0}(E_{\rm F})$ is the DOS at the Fermi level. 
We define $x\equiv q/2k^{0}_{\rm F}$ where $k^{0}_{\rm F}$ is the Fermi wave number. The values of the free electron gas are applied to $D_{0}(E_{\rm F})$ and $k^{0}_{\rm F}$. 
$\Pi_0(x)$ is the three-dimensional Lindhardt function, 
\begin{eqnarray}
\Pi_0(x)=\frac{1}{2}\bigg[1+\frac{1-x^2}{2x}\ln \bigg| \frac{1+x}{1-x} \bigg|\bigg]. 
\end{eqnarray}
Since we start with the divalent state for Yb ($Z^{\rm Yb}=+2$), the free electron gas consists of nine electrons from three Al$^{3+}$ ions and two electrons from one Yb$^{2+}$ ion. 
The potentials in the Fourier space are shown in Fig. \ref{potential}. 
Both figures exhibit explicitly that the amplitude of each potential is strongly reduced by the screening effect, in particular, in the small $|{\mathbf q}|$ region. 

Finally, the pseudo-potential for the conduction electrons in the NFE method for the whole lattice is given as follows 
\begin{eqnarray}
\tilde{V}_{\mathbf G}=\tilde{v}^{\rm Yb}_{\mathbf G}+\tilde{v}^{\rm Al}_{\mathbf G} \sum_{\pmb{\tau}}{\rm e}^{-{\rm i}\mathbf G \cdot \pmb{\tau}}. 
\end{eqnarray}
Thus, the Hamiltonian for the conduction electrons with spin $\sigma$ is written as
\begin{eqnarray}
{\cal H}^{\rm c}_{\sigma}=\sum_{\mathbf k}
\left(
\begin{array}{cccc}
    \epsilon_{\mathbf k_1}+\tilde{V}_{0} 
  & \tilde{V}_{\mathbf G_1-\mathbf G_2}
  & \ldots & \tilde{V}_{\mathbf G_1-\mathbf G_n}\\
    
    \tilde{V}_{\mathbf G_1-\mathbf G_2}^{*} 
  & \epsilon_{\mathbf k_2}+\tilde{V}_{0}
  & \ldots & \tilde{V}_{\mathbf G_2-\mathbf G_n}\\
  
    \vdots 
  & \vdots 
  & \ddots 
  & \vdots\\
  
    \tilde{V}_{\mathbf G_1-\mathbf G_n}^{*} 
  & \tilde{V}_{\mathbf G_2-\mathbf G_n}^{*} 
  & \ldots & \epsilon_{\mathbf k_n}+\tilde{V}_{0}
\end{array}
\right). 
\label{eqn:hc}
\end{eqnarray}
$\mathbf G_i$ and $\mathbf k_i$ with an integer set $i\equiv (i_1,i_2,i_3)$ are defined as
\begin{eqnarray}
\mathbf G_i&\equiv &i_1 \mathbf G_0^1 +i_2 \mathbf G_0^2 +i_3 \mathbf G_0^3 ,
\label{eqn:recg}\\
\mathbf k_i&\equiv & \mathbf k+\mathbf G_i ,
\end{eqnarray}
where $\mathbf G_0^\mu$ ($\mu=1,2,3$) denotes the primitive translation vector of the reciprocal lattice. 
Note here that ${\cal H}^{\rm c}_{\sigma}$ is independent of the spin index $\sigma$. 

Because of the strong spin-orbit coupling of f-electrons, the local f-electron states are specified by the total angular momentum $J$ and its $z$-component $M$.  For Yb ion, the local f-electron states consist of $J=7/2$ and $J=5/2$ states. While the former has the eight-fold degeneracy and is located close to the Fermi level, the latter has the six-fold degeneracy and is far from the Fermi level for YbAl$_3$~\cite{hari,lee03}. The energy difference between the $J=7/2$ and $J=5/2$ states is defined as $\Delta$, which is of the order of 0.1 Ry~\cite{hari,lee03}. Although the $J=5/2$ state is almost irrelevant to the low-energy physics, we include this state in ${\cal H}^{\rm f}$. The localized f term of the Hamiltonian is written as 
\begin{eqnarray}
{\cal H}^{\rm f}=\sum_{i, J, M}E_{\rm f}^{J} f^{\dag}_{i JM}f_{i JM},
\end{eqnarray}
where $f^{\dag}_{i JM}$ ($f_{i JM}$) is a creation (annihilation) operator of a localized f-electron with the site index $i$, and the $z$-component $M$ of the total angular momentum $J=7/2$ and $5/2$. The 4f-energy level $E_{\rm f}^{J}$ is a tuning parameter in the present scheme. 
Note here that although it is easy to take account of the crystal field effects it is not discussed in the present study, for simplicity.

Next, let us consider the matrix element of the hybridization between the conduction electron and the f-electron. The plane wave and the localized f-electron can be expanded by the spherical harmonics, respectively. Considering the orthogonality of the spherical harmonics, the matrix element of the hybridization term is written as
\begin{eqnarray}
V_{\mathbf k\sigma JM } 
&\hspace{-2mm}\equiv &\hspace{-2mm}\langle \mathbf k \sigma | {\cal H}^{\rm hyb} |f_{\mathbf k JM} \rangle \nonumber \\
&\hspace{-2mm}=&\hspace{-2mm}\sum_{m}a^{JM}_{lm\sigma}Y_{lm}(\theta_{\mathbf k},\phi_{\mathbf k}) V_{knl},
\label{eqn:hybcf}
\\
V_{knl}&\hspace{-2mm}\equiv & \hspace{-2mm}\frac{4\pi (-1)^l}{\sqrt{\Omega_0}}\int_{0}^{\infty}
{\rm d}r \,\,r^2 R_{nl}(r)v(r)j_l(kr), 
\end{eqnarray}
where $\sigma$ stands for the spin index of the conduction electron. $|f_{\mathbf k JM}\rangle$ denotes the Fourier transformation of $|f_{i JM}\rangle$. $Y_{lm}(\theta_{\mathbf k}, \phi_{\mathbf k})$ is the spherical harmonics with the orbital angular momentum $l$ and its $z$-component $m$ where $\theta_{\mathbf k}$ and $\phi_{\mathbf k}$ denote the direction of the wave vector $\mathbf k$. $j_l(kr)$ is the spherical Bessel function.~\cite{hanz87,kont96}
For simplicity, we regard $V_{knl}$ as a constant parameter $V$. Although $V_{\mathbf k \sigma JM }$ is independent of the radial part of the wave vector $\mathbf k$, the angular dependence of $\mathbf k$ still remains. $a^{JM}_{lm\sigma}$ is the Clebsch-Gordan coefficient. For the f-electrons with $l=3$, $J=7/2$ and $5/2$, it is given by
\begin{eqnarray}
a^{JM}_{lm\sigma}=
\left \{
\begin{array}{cl}
\sqrt{\frac{1}{7}\left(\frac{7}{2}+M\sigma \right)}\delta_{m,M-\frac{\sigma}{2}}&\left(J=\frac{7}{2}\right)\\
-\sigma \sqrt{\frac{1}{7}\left(\frac{7}{2}-M\sigma \right)}\delta_{m,M-\frac{\sigma}{2}}&\left(J=\frac{5}{2}\right), 
\end{array}
\right. 
\label{eqn:clbgld72}
\end{eqnarray}
where $\sigma$ stands for $1$ ($\uparrow$ spin) and $-1$ ($\downarrow$ spin), respectively. 

Therefore, substituting eq. (\ref{eqn:clbgld72}) for eq. (\ref{eqn:hybcf}), $V_{\mathbf k \sigma J M }$ is written as~\cite{hanz87} 
\begin{eqnarray}
V_{\mathbf k \sigma J M }\hspace{-1mm}=
\hspace{-1mm}\left \{
\begin{array}{cl}
\hspace{-2mm}\sqrt{\frac{1}{7}\left(\frac{7}{2}+M\sigma \right)}Y_{3,M-\frac{\sigma}{2}}(\theta_{\mathbf k},\phi_{\mathbf k}) V & \hspace{-3mm} \left(J=\frac{7}{2} \right) \\
\hspace{-2mm}-\sigma \sqrt{\frac{1}{7}\left(\frac{7}{2}-M\sigma \right)}Y_{3,M-\frac{\sigma}{2}}(\theta_{\mathbf k},\phi_{\mathbf k}) V & \hspace{-3mm} \left(J=\frac{5}{2}\right). 
\end{array}
\right. 
\end{eqnarray}

Finally, the effective Hamiltonian for the band energy is written as
\begin{eqnarray}
{\cal H}^{\rm band}=
\left (
\begin{array}{cc|c}
{\cal H}^{\rm c}_{\uparrow} 
& 0 
&\vspace{-2mm}
\\

&
&{\cal H}^{\rm hyb} 
\\

0 
& {\cal H}^{\rm c}_{\downarrow} 
&  \\ \hline

{\cal H}^{\rm hyb*}\hspace{-10mm}
& 
& {\cal H}^{\rm f}
\end{array}
\right). 
\end{eqnarray}
Summarizing this section, the effective Hamiltonian for the band energy is constructed by the combination of the NFE scheme for the conduction electrons and the hybridization with the localized f-electrons. This Hamiltonian includes only a few parameters (the hybridization $V$ and the 4f-level $E_{\rm f}^{J}$ in addition to $Z^{\alpha}$ and $r^{\alpha}_{\rm c}$), which can be estimated from the LDA and experimental results, as will be is discussed in the next section.

\section{Results}
\label{results}
Although all reciprocal lattice vectors should be included in eq. (\ref{eqn:hc}) in the NFE method, the present calculations actually converge with $i_\mu$=$-3$ to $+3$ ($\mu=1,2,3$) in eqs. (\ref{eqn:hc}) and (\ref{eqn:recg}). 
Then, the size of the Hamiltonian matrix ${\cal H}^{\rm band}$ becomes $700$ ($=7^3 \times 2 +14$) for each wave vector $\mathbf k$. Diagonalizing this matrix numerically, we can obtain all eigenvalues for each $\mathbf k$. Although we divide the first Brillouin zone into $N=120^3$ pieces to perform $\mathbf k$-integration, the calculated physical quantities almost converge at $N=60^3$. 
The actual calculations are carried out within 1/48 of the first Brillouin Zone. 
In summary, we use the following parameters in the present study, the energy difference between the $J=7/2$ and $J=5/2$ levels are chosen as $\Delta \equiv E_{\rm f}^{7/2}-E_{\rm f}^{5/2}=0.1$ Ry. 
The pseudo-potential parameters are 
$r^{\rm Al}_{\rm c}=0.61$ \r{A} and $r^{\rm Yb}_{\rm c}=2.0$ \r{A}. The valences of the ions are simply assumed as $Z^{\rm Yb}=+2$ and $Z^{\rm Al}=+3$, respectively. 
The Yb 4f $J=7/2$ level $E_{\rm f}^{7/2}$ and the hybridization amplitude $V$ are the main tuning parameters. 
Except the result of the optical conductivity, we discuss the physical properties at absolute zero temperature. 

\begin{figure}[tb]
\begin{center}
\includegraphics[width=7cm]{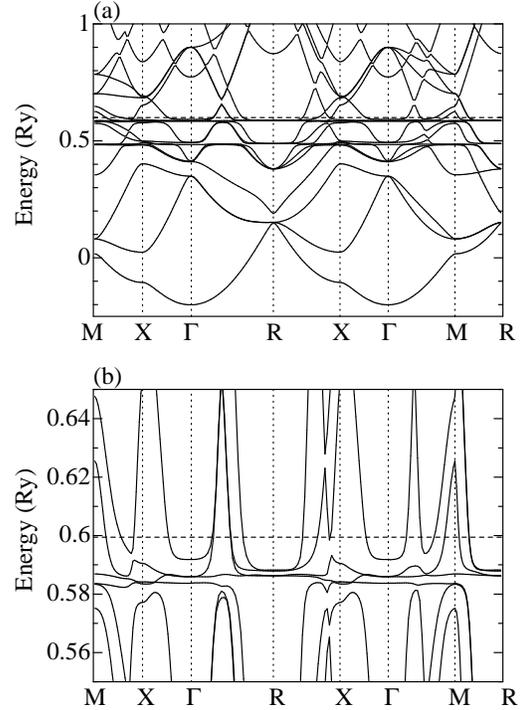}
\end{center}
\caption{
Energy band dispersions of YbAl$_3$ for $V=0.03$ Ry and $E_{\rm f}^{7/2}=0.59$ Ry. The horizontal dashed-line stands for the Fermi level. The upper panel (a) shows the overall structure and the lower panel (b) is the extreme magnification close to the Fermi level ($E_{\rm F}\sim 0.599$ Ry). 
}
\label{band}
\end{figure}

Figure \ref{band} shows the band dispersions of YbAl$_3$ obtained by the present scheme for $E_{\rm f}^{7/2}=0.59$ Ry and $V=0.03$ Ry. 
Note here that the overall band dispersion is shifted by 0.48 Ry in order to adjust the origin of our band to that of the LDA band (Fig. \ref{ldaband}).  
The almost flat bands at $\sim 0.59$ Ry and at $\sim 0.49$ Ry correspond to the Yb 4f $J=7/2$ and $J=5/2$ levels, respectively. 
However, because of the hybridization term, the 4f-bands become dispersive, which is shown in the lower panel of Fig. \ref{band} for the $J=7/2$ band. The Fermi level is located at a little above the $J=7/2$ level and intersects some dispersive bands, which explicitly indicates that the obtained result describes the metallic state of YbAl$_3$. 
Furthermore, we stress that the hybridization gap does not appear within the present treatment of the c-f hybridization and the parameter sets. 

It seems that although the details are different from that obtained by LDA, the overall structure around and below the Fermi level corresponds well to the LDA result (Fig. \ref{ldaband}(a)). The physical properties mainly depend on the low-energy electronic structure. 
It is consistent with the LDA result (Fig. \ref{ldaband}(b)) except that close to M point. 
There are small closed hole-like Fermi surfaces on $\Gamma$-R line and closed electron-like Fermi surfaces are located around $\Gamma$ and R points, which are consistent with those of the LDA band. However, the neck at M point in the LDA band (Fig. \ref{ldaband}(b)) is not reproduced. 

The band structures are calculated with other values of $V$ in order to consider the robustness of the Fermi surface shapes. Although the $J=7/2$ band becomes dispersive with an increase of the amplitude of the hybridization, it seems that there is no topological change of the Fermi surface for $E_{\rm f}^{7/2} \leq 0.64$ Ry. It is because the main $J=7/2$ band does not directly intersect the Fermi level. 

\begin{figure}[tb]
\begin{center}
\includegraphics[width=7cm]{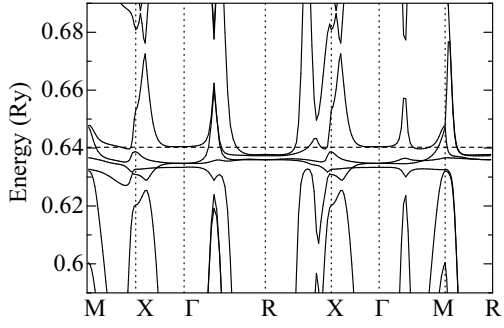}
\end{center}
\caption{Band structure around the Fermi level (dashed line) for $V=0.03$ Ry and $E_{\rm f}^{7/2}=0.64$ Ry. 
}
\label{band12}
\end{figure}

With an increase of $E_{\rm f}^{7/2}$ level to higher energy region, it is approaching to the Fermi level. For $E_{\rm f}^{7/2} \geq 0.64$ Ry, the electron-like Fermi surface around $\Gamma$ point vanishes, which is shown in Fig. \ref{band12}. 
This band structure is in partial agreement with the LDA band with the upward-shift of the 4f-levels (Fig. \ref{ldaband}(c)). Note that the hybridization gap does not appear in this case, either. 

\begin{figure}[b]
\begin{center}
\includegraphics[width=7cm]{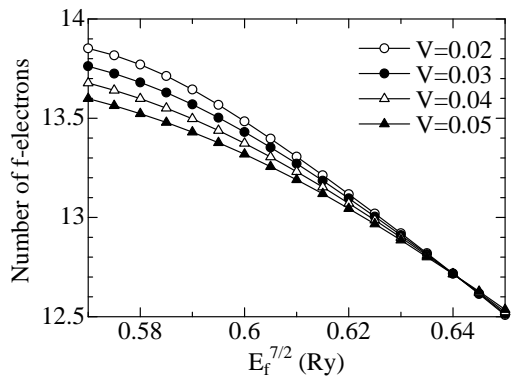}
\end{center}
\caption{
Number of f-electrons as a function of $J=7/$2 level $E_{\rm f}^{7/2}$ for several choices of hybridization $V$. 
}
\label{particle}
\end{figure}

The number of f-electrons is given by 
\begin{eqnarray}
n^{\rm f}=\frac{1}{N}\sum_{\mathbf k \alpha JM}|u^{JM}_{\mathbf k \alpha}|^2\,\theta (E_{\rm F}-E_{\mathbf k \alpha}), 
\end{eqnarray}
where $E_{\mathbf k \alpha}$ and $u^{JM}_{\mathbf k \alpha}$ are the eigenvalue and eigenvector of ${\cal H}^{\rm band}$ with the wave vector $\mathbf k$ and band index $\alpha$. 
The results for several choices of the hybridization $V$ are shown in Fig. \ref{particle}. 
With an increase of the $J=7/2$ level $E_{\rm f}^{7/2}$ and/or the hybridization $V$, the deviation from $f^{14}$ configuration becomes conspicuous.  
Compared with the experimental mean-values of Yb valence number (2.65 $\sim$ 2.8), $E_{\rm f}^{7/2}$ is expected as 0.59 $\sim$ 0.60 Ry. 
While the band structure in Fig. \ref{band12} can partially reproduce the LDA result with the upward-shift of the 4f-levels (Fig. \ref{ldaband}(c)), the number of f-electrons becomes too small ($n_{\rm f} \sim 12.72$ $<$ 13). 
Thus, we hereafter focus on the parameter set in Fig. \ref{band} ($n_{\rm f} \sim 13.57$ for $E_{\rm f}^{7/2}=0.59$ Ry and $V=0.03$ Ry). 

\begin{figure}[tb]
\begin{center}
\includegraphics[width=7cm]{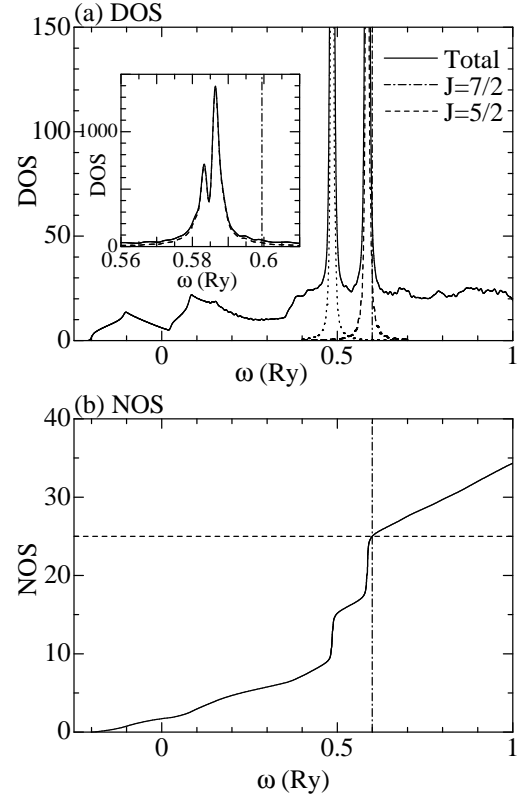}
\end{center}
\caption{(a) DOS for $V=0.03$ Ry and $E_{\rm f}^{7/2}=0.59$ Ry. The solid, dashed and dotted lines stand for total, $J=7/2$ and $J=5/2$ DOS's, respectively. The inset shows the magnification around the $J=7/2$ level. The panel (b) shows NOS where the dashed line stands for the total particle number of YbAl$_3$ per unit cell, which corresponds to 25. The dash-dotted line represents the Fermi level in both panels. 
}
\label{dnos}
\end{figure}

Next, let us study the property of the DOS and the number of states (NOS). Figure \ref{dnos} shows the total DOS, the partial f-DOS and the NOS, which are defined as 
\begin{eqnarray}
D^{\rm tot}(\omega)&=&\frac{1}{N}\sum_{\mathbf k \alpha}\delta (\omega-E_{\mathbf k \alpha}), \\
D^{\rm f}_{J}(\omega)&=&\frac{1}{N}\sum_{\mathbf k \alpha M}|u^{JM}_{\mathbf k \alpha}|^2\delta (\omega-E_{\mathbf k \alpha}), \\
N(\omega)&=&\int_{-\infty}^{\omega} {\rm d}\omega' D^{\rm tot}(\omega'). 
\end{eqnarray}

The overall structure of DOS is roughly similar to the result of ref. \ref{bib:lee03}. Two sharp peaks appear around the $J=7/2$ and $J=5/2$ levels because of the strongly localized character of the f-electrons. However, both peaks are broadened due to the c-f hybridization. 
With an increase of the hybridization amplitude, the peak around the $J=7/2$ level splits into two parts, which is shown in the inset of Fig. \ref{dnos}(a). The energy difference between those peaks is about 0.005 Ry for the given parameters. It is important that the hybridization gap does not explicitly appear because of the angular dependence of $\mathbf k$ in the c-f hybridization, the periodic potential and the band-folding into the first Brillouin zone. 
Of course, although the further increase of the hybridization amplitude gives rise to the growth of the pseudo-gap, an explicit gap never appears within the range of the investigated parameters. 
Since the partial 4f-DOS has the tail due to the c-f hybridization, the Fermi level partially intersects the $J=7/2$ bands. 
Thus, the number of f-electrons per unit cell deviates from integer ($n^{\rm f}=14$) except the $V=0$ case. 
Figure \ref{dnos}(b) shows the NOS for reference, in which the jumps appear around the $J=7/2$ and $J=5/2$ levels. The presence of the c-f hybridization smears the jumps. 
Similar to the band dispersion, the result of the DOS also indicates that YbAl$_3$ is in a valence fluctuating state, which is in agreement with the experimental results. 

The Sommerfeld coefficient $\gamma$ is written as
\begin{eqnarray}
\gamma =\frac{\pi^2}{3} k_{\rm B}^2 D^{\rm tot}(E_{\rm F}) N_{\rm A}, 
\end{eqnarray}
where $N_{\rm A}$ is Avogadro's number. For $E_{\rm f}^{7/2}=0.59$ Ry and $V=0.03$ Ry, $\gamma$ is about 8.55 mJ$\cdot$ K$^{-2}$$\cdot$mol$^{-1}$. For other values of the parameters, $\gamma$ is in the range from 5 to 10 mJ$\cdot$ K$^{-2}$$\cdot$mol$^{-1}$ . These values are smaller than the experimental value ($\sim 40$ mJ$\cdot$ K$^{-2}$$\cdot$mol$^{-1}$) since the effect of electron correlations between the f-electrons is not taken into account, which will be discussed later. 

Finally, let us investigate the optical conductivity $\sigma(\omega)$ and its temperature dependence. By using Kubo formula, the contribution of the direct transition, except Drude part, is given by
\begin{eqnarray}
{\rm Re} \sigma^{\rm direct}(\omega)
\hspace{-4mm}
&=&\hspace{-4mm}\frac{\pi}{\omega N}
\sum_{\mathbf k \alpha \ne \beta} 
\left( j_{x}^{\alpha \beta}(\mathbf k) \right)^2
\left\{ f(E_{\mathbf k \alpha})-f(E_{\mathbf k \beta}) \right\}\nonumber \\
\hspace{-4mm}&\times &\hspace{-4mm}
\delta \left(\omega-(E_{\mathbf k \beta}-E_{\mathbf k \alpha}) \right), 
\label{eqn:sigma}
\end{eqnarray}
where $j_{x}^{\alpha \beta}(\mathbf k)$ stands for the current matrix element between the band index $\alpha$ and $\beta$. $f(E)$ is the Fermi distribution function. For simplicity, we neglect the wave vector and band index dependences of the current term in the present study. Namely, we set $j_{x}^{\alpha \beta}(\mathbf k)={\rm constant}$. 

\begin{figure}[tb]
\begin{center}
\includegraphics[width=7cm]{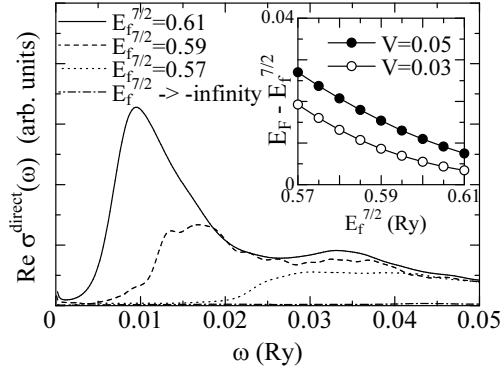}
\end{center}
\caption{Optical conductivity for a few choices of $J=7/2$ level and $V=0.03$ Ry at absolute zero temperature. The dash-dotted line stands for the result for $E_{\rm f}^{7/2}\rightarrow -\infty$ and $V=0$. Drude parts are not shown. The inset shows the energy difference between the Fermi level $E_{\rm F}$ and the $J=7/2$ level $E_{\rm f}^{7/2}$. 
}
\label{sigma_ef}
\end{figure}

Figure \ref{sigma_ef} shows the conductivity for a few choices of $E_{\rm f}^{7/2}$ at absolute zero temperature, in which Drude parts are not included. To consider contributions of the f-electrons, we also calculate the conductivity for $E_{\rm f}^{7/2}\rightarrow -\infty$ and $V=0$. 
Except the $E_{\rm f}^{7/2}\rightarrow -\infty$ case, the peaks appear at low energy region ($\omega \sim$ 0.01 to 0.03 Ry, which corresponds to the range from 0.15 to 0.4 eV). With a decrease of the $E_{\rm f}^{7/2}$ level, the peak structures are smeared and the peak positions shift to the high energy region. For $E_{\rm f}^{7/2}\rightarrow -\infty$, the amplitude is extremely smaller than those of other values of $E_{\rm f}^{7/2}$. Thus, we can conclude that these peaks at $\omega \sim$ 0.01 to 0.03 Ry are mainly due to the excitations from the $J=7/2$ bands to the conduction bands above $E_{\rm F}$ with the same $\mathbf k$, 
in which the peak position for $E_{\rm f}^{7/2}=0.59$ Ry corresponds roughly to the energy difference between the center of gravity position of the $J=7/2$ DOS and the Fermi level (shown in the inset of Fig. \ref{dnos}(a)). 
The peak due to the excitations from the $J=5/2$ bands appears at $\omega \sim$ 0.1 Ry. 
Note that we replace the delta function with a gaussian function in eq. (\ref{eqn:sigma}) on the numerical calculations, so that this broadening effect gives rise to the sharp peaks close to $\omega = 0$. With a decrease of a half width of the gaussian function, the peak amplitude becomes small. 

The optical conductivity observed in the experiment shows the pseudo-gap structure at very low energy region and the mid-infrared peak around $\hbar \omega =0.25$ eV~\cite{okam04}. The origin of the mid-infrared peak has been attributed to the excitations across the c-f hybridization gap, seen in the simple PAM~\cite{okam04}. 
However, in our model, the hybridization gap does not exist (shown in Fig. \ref{band} and the inset of Fig. \ref{dnos}(a)).

The structure of the calculated conductivity is consistent with that of experimental result. The peaks found in the range from 0.01 to 0.03 Ry seem to correspond to the mid-infrared ones observed in the experiment.~\cite{okam04} The energy difference between the split peaks in the $J=7/2$ DOS (the inset of Fig. \ref{dnos}) is at most 0.005 Ry ($\sim$ 0.07 eV), which is too small for the energy scale of the observed mid-infrared peak. 
Of course, with an increase of the hybridization amplitude, the energy difference between the split peaks in the $J=7/2$ DOS is enlarged and the tail in the partial f-DOS grows. In this case, since the number of f-electrons has to be in the range from 13 to 14, the Fermi level shifts to the higher energy region away from the $J=7/2$ level (shown in the inset of Fig. \ref{sigma_ef}). Therefore, the peak position of the conductivity also shifts to the higher energy region. Thus, we conclude that the origin of the mid-infrared peak is not the excitations between the split bands due to the c-f hybridization but the excitations from the $J=7/2$ bands to the conduction bands above $E_{\rm F}$ with the same $\mathbf k$. 

\begin{figure}[tb]
\begin{center}
\includegraphics[width=7cm]{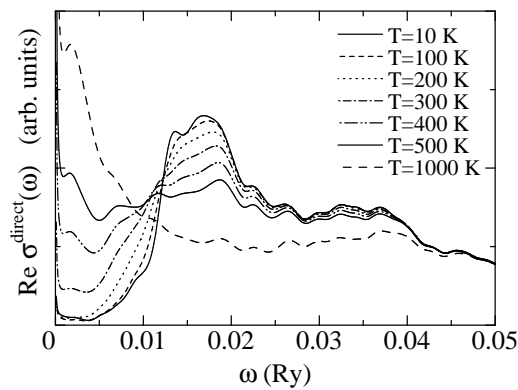}
\end{center}
\caption{Temperature dependence of optical conductivity for $V=0.03$ Ry and $E_{\rm f}^{7/2}=0.59$ Ry. 
}
\label{sigma_temp}
\end{figure}

The temperature dependence of the conductivity is shown in Fig. \ref{sigma_temp}. Drude parts are not included. 
The pseudo-gap structure at low energy region is gradually filled with an increase of temperature. The characteristic temperature for disappearance of the pseudo-gap is about 500 K for the present parameters. 
If correlation effects between the f-electrons are included, a Drude peak with a finite width will be added at finite temperatures, and the width will become larger with an increase of temperature. Filling of the pseudo-gap will become faster than that of the non-interacting case since the DOS also depends on temperature.~\cite{saso04} 

\section{Summary and Discussions}
\label{sumdis}
We have constructed the effective model of the band energy for heavy fermion compound YbAl$_3$. 
In order to take account of the effect of the periodic potential reflecting the crystal lattice structure, the NFE method with the empty-core pseudo-potential is employed for the conduction electrons, in which the screening effect is considered by using the dielectric function of the free electron gas. The localized bases are employed for the 4f-electrons because of the strongly localized character. Furthermore, the hybridization between the NFE conduction band and the f-band is accurately taken into account. 
Adjusting only a few parameters and diagonalizing the obtained Hamiltonian, we have calculated the band dispersions, the DOS, the Sommerfeld coefficient and the optical conductivity. 

While the often-used simple PAM and its extensions usually yield a hybridization gap, {\it our realistic model explicitly denies the existence of the hybridization gap in the metallic system.} 
The number of f-electrons deviates from integer and covers the experimentally observed range. Thus, the valence fluctuating state is reproduced. The obtained band structure and the DOS are mostly consistent with the LDA results. Furthermore, we find that the origin of the experimentally observed mid-infrared peak is due to the excitations from the dispersive 4f $J=7/2$ bands to the conduction bands above $E_{\rm F}$ with the same wave vector $\mathbf k$, and not the transition across the hybridization gap. 
We conclude that the present effective model with only a few parameters may describe the physical properties of YbAl$_3$. 
However, by varying the parameters, we can only partially reproduce the LDA band with the upward-shift of the 4f-levels which is in agreement with the dHvA result. In addition, the number of 4f-electrons becomes unfortunately too small. This discrepancy in the present approach is to be improved in a future. 
Nevertheless, the present approach seems to be applicable to a wide range of heavy electron metals, and our conclusion that the hybridization gap is absent would be true for most cases. 

Finally, we have to discuss correlation effects of the 4f-electrons. 
Although it is well known that the rare-earth compounds are strongly correlated electron systems, we focus on the construction of the effective Hamiltonian of the band energy in the present study. 
Correlation effects should be taken into account in order to estimate physical quantities accurately, which is now under investigation. 
However, the Coulomb interactions are strongly screened at low energy region in heavy fermion systems, so that the Fermi liquid behavior is often realized at low temperature region. For $T<35$ K, YbAl$_3$ shows the Fermi liquid property. 
Thus, if the 4f-levels and the c-f hybridization amplitude employed in the present study are regarded as the renormalized quantities, correlation effects may be taken into account partially. 
Of course, the physical properties caused by dynamical quantum fluctuations are not included. 
Our model may be a good starting point to systematically consider such correlation effects. 


\section*{Acknowledgement}
The authors would like to thank H. Harima for providing us with Figs. \ref{ldaband}(a)-(c) and valuable discussions. They are indebted to H. Okamura for illuminating discussions. 
This research was partially supported by the Ministry of Education, Science, Sports and Culture, Japan. 

\end{document}